\documentclass[aps,pra,showpacs,twocolumn,superscriptaddress]{revtex4}

\usepackage{graphicx}
\usepackage{dcolumn}
\usepackage{bm}
\usepackage{color}
\usepackage[latin1]{inputenc}
\usepackage{ulem}

\begin{document}
\title{Protecting the $\sqrt{\rm SWAP}$ operation from general and residual errors\\ by continuous dynamical decoupling}
\author{F. F. Fanchini}
 \email{fanchini@fc.unesp.br}
\affiliation{Faculdade de Ci\^encias, UNESP - Universidade Estadual Paulista, Bauru, SP, 17033-360, Brazil}
\author{R. d. J. Napolitano}
\affiliation{Instituto de F\'{\i}sica de S\~{a}o Carlos,
Universidade de S\~{a}o Paulo, P.O. Box 369, S\~{a}o Carlos, SP, 13560-970, Brazil}
\author{B. \c{C}akmak}
\affiliation{Instituto de F\'{\i}sica Gleb Wataghin,  Universidade
Estadual de Campinas, P.O. Box 6165, Campinas, SP, 13083-970, Brazil}
\author{A. O. Caldeira}
\affiliation{Instituto de F\'{\i}sica Gleb Wataghin,  Universidade
Estadual de Campinas, P.O. Box 6165, Campinas, SP, 13083-970, Brazil}
\date{\today}

\begin{abstract}
We study the  occurrence of errors in a continuously decoupled two-qubit state during a $\sqrt{\rm SWAP}$ quantum operation under decoherence. We consider a realization of this quantum gate based on the Heisenberg exchange interaction, which alone suffices for achieving universal quantum computation. Furthermore, we introduce a continuous-dynamical-decoupling scheme that commutes with the Heisenberg Hamiltonian to protect it from the amplitude damping and dephasing errors caused by the system-environment interaction. We consider two error-protection settings. One protects the qubits from both amplitude damping and dephasing errors. The other features the amplitude damping as a residual error and protects the qubits from dephasing errors only. In both settings, we investigate the interaction of qubits with common and independent environments separately. We study how errors affect the entanglement and fidelity for different environmental spectral densities.

\end{abstract}

\maketitle

\section{INTRODUCTION}
Quantum computers use superposition and entanglement of qubits to outperform digital computers \cite{deutsch92,shor}. The advent of these machines will unquestionably encompass a radical transformation in the way we simulate quantum-mechanical processes \cite{feynman82}, imparting a plethora of new achievements in science and technology. However, to take advantage of the benefits of reliable quantum information processing, we depend on the development of efficient ways to avoid or recover from the errors that, induced by environmental interaction \cite{zurek03}, occur in the state of our quantum system.

Accordingly, several strategies to protect quantum information, particularly during the operation of a quantum gate, have been designed, including quantum error correcting codes \cite{qecc}, fault-tolerant quantum computing \cite{ftqc}, decoherence-free subspaces \cite{dfs,lidar}, etc. One of the most effective methods to protect the state of a quantum system from decoherence is called dynamical decoupling (DD), which has been extensively studied in the literature, both theoretically \cite{ddtheory,viola99,zanardi99,byrd02,facchi05,khodjasteh08,khodjasteh09,khodjasteh09b,khodjasteh10,lidar14} and experimentally \cite{ddexperiment,du09,damodarakurup09,lange10,souza11,naydenov,sar12}, and is the main subject we focus in this paper. The DD approach is based on applying a sequence of external control pulses to the quantum system to be protected, in order to suppress the errors arising from its coupling with the environment.  In other words, we introduce an additional Hamiltonian, called the control Hamiltonian, that acts on the Hilbert space of the system, averaging out the effects of the environmental perturbations.

Alternatively, instead of control pulses, it is also possible to apply continuous external fields to decouple the system from the environmental interactions. This scheme, known as continuous dynamical decoupling (CDD), has attracted a lot of attention in recent years \cite{contdd,fonsecaromero05,clausen10,xu12,fanchini022329,fanchini07,cai,laraoui,aiello,mkhitaryan}. The CDD procedure is more experimentally friendly than pulsed procedures and it also sets a natural stage for the implementation of two-qubit quantum gates \cite{pleniogates,timoney11,bermudez12}. For instance, nitrogen vacancy (NV) centers in diamonds have recently been shown to be strong candidates for use in the field of quantum technologies, with possible applications ranging from biological systems to quantum computing protocols \cite{nvcenter}. Furthermore, for reducing the damage of environmental perturbations in NV centers, the CDD procedure has again proven to be very important \cite{pleniodiamond}.

In this work, we consider CDD of a two-qubit system going through a $\sqrt{\rm SWAP}$ operation while interacting with a bosonic environment. While Refs. \cite{pleniogates,timoney11,bermudez12} treat the case of magnetic noise, where the couplings between the qubits and their environment are orders of magnitude stronger than the interqubit interaction, here we consider the case in which the two-qubit gate interaction is stronger than the perturbations by the environment. We obtain a simple control prescription which allows us to prove the effectiveness of our method in a realistic decoherence model. The effect of the environment is simulated by two different quantum channels -- amplitude damping and dephasing -- simultaneously and independently coupled to the qubits with different coupling strengths. We also study the effects of residual errors when the CDD protection is supplied just against the predominant error source, i.e., the one with the strongest coupling. In the present context, we consider the residual error as arising from the amplitude damping channel, while the qubits are dynamically decoupled from dephasing. We adopt the concurrence \cite{concurrence} and the fidelity as the figures of merit of the CDD procedure. We show that the adopted CDD scheme provides nearly full protection against environmental effects when both error mechanisms are present and, for a residual amplitude-damping environment, a super-ohmic spectral density of states is more destructive than an ohmic one. Furthermore, in the absence of the CDD, we see that, in the case of a common environment for the qubits, both entanglement and fidelity decay more slowly than in the case of independent environments.

The present paper is organized as follows. Section II reviews the CDD procedure for the case of a $\sqrt{\rm SWAP}$ quantum gate. Section III describes our model to simulate the error sources. Section IV shows the results for the CDD protection of the $\sqrt{\rm SWAP}$ from both amplitude damping and dephasing, and from dephasing only, when amplitude damping is treated as a residual error channel. Finally, a conclusion is presented in Sec. V. { The detailed solution of the master equation, that gives the dynamics of the reduced density matrix of the system, is presented in the Appendix.}

\section{The model and continous dynamical decoupling}
To illustrate our protective scheme, we begin by assuming
that the interaction Hamiltonian between the qubit system and the rest of the universe is of the form
\begin{eqnarray}
H_{\rm int}={\bf B}^{(1)}\cdot {\bm \sigma}^{(1)}+{\bf B}^{(2)}\cdot {\bm \sigma}^{(2)},\label{Hint}
\end{eqnarray}
where ${\bf B}^{(s)}=\sum _{m=1} ^{3}B_{m}^{(s)}{\bf
\hat{x}}_{m}$, for $s=1,2$,  with ${\bf \hat{x}}_{1}\equiv{\bf
\hat{x}}$, ${\bf \hat{x}}_{2}\equiv{\bf \hat{y}}$, ${\bf
\hat{x}}_{3}\equiv{\bf \hat{z}}$, and $B_{m}^{(s)}$, for $s=1,2$
and $m=1,2,3$, are Hermitian operators that act on the
environmental Hilbert space. The main approach of the
dynamical decoupling method \cite{ddtheory,viola99,zanardi99,byrd02,facchi05,khodjasteh08,khodjasteh09,khodjasteh09b,khodjasteh10,lidar14} in order to reduce errors
on the system, is to eliminate
the effect of the interaction Hamiltonian by an external
control Hamiltonian. Mathematically, such a condition can
be written as
\begin{equation}
\int ^{t_{c}}_{0} U^{\dagger}_{c}(t)H_{\rm int}U_{c}(t) dt=0,\label{condition}
\end{equation}
where $U_c(t)$ is the time evolution operator associated with the
control Hamiltonian $H_c$ and $t_{c}=2\pi /\omega $. Equation (\ref{condition}) results from a Magnus expansion \cite{magnus} used to describe the total interaction-picture evolution operator, in the limit in which $t_{c}\rightarrow 0$. In this limit, only the first term of the expansion, given by the integral in Eq. (\ref{condition}), is, in general, non-zero. Thus, in this ideal circumstance, by imposing Eq. (\ref{condition}) we are ensuring the complete elimination of the perturbing interactions. Here, however, we use Eq. (\ref{condition}) as a mere guide to develop the present approach, for our focus is on the realistic situation of a finite $t_{c}$. In the present paper, we will gauge numerically the efficacy of the resulting approximate method.

In order to control the intensity of
the exchange interaction, possible candidates for the physical
qubits should be, for example, properly built tunable charge
qubits \cite{schon}. Although, in this particular case, physical
reasoning leads us to assume that each qubit is coupled to its own
environment, we shall, for the sake of completeness, also study
the case of a common environment. For our present purposes, we
assume that the particular form of Eq. (\ref{Hint}) is, in the case of a common environment, given by
\begin{equation}
H_{int}=\left({\bm \sigma}^{(1)} + {\bm \sigma}^{(2)}\right)\cdot
\left({\bm \lambda}B + {\bm\lambda }^\ast
B^\dagger\right),\label{collective}
\end{equation}
where
${\bm B}^{(s)}={\bm \lambda}B+{\bm \lambda}^{\ast}B^{\dagger}$,
for $s=1,2$, ${\bm \lambda}$ is an arbitrary complex
three-dimensional  vector, and $B$ is a scalar operator that acts
on the environmental Hilbert space. However, when the qubits are physically located
sufficiently far apart, as for tunable charge
qubits \cite{schon}, it is reasonable to suppose that their
individual surroundings act as uncorrelated, independent
environments. In that case, the particular form we assume for Eq.
(\ref{Hint}) is written as
\begin{eqnarray}
H_{int}&=&{\bm \sigma}^{(1)}\cdot\left({\bm \lambda^{(1)}}B^{(1)} +
{\bm\lambda^{(1)}}^\ast {B^{(1)}}^\dagger\right)\nonumber\\
&+&{\bm \sigma}^{(2)}\cdot\left({\bm \lambda^{(2)}}B^{(2)} +
{\bm\lambda^{(2)}}^\ast {B^{(2)}}^\dagger\right)\label{independent}
\end{eqnarray}
where $B^{(s)}$, for $s=1,2$, acts on the environmental Hilbert
space  of the $s$-th qubit, and ${\bm \lambda^{(s)}}$ is an
arbitrary complex three-dimensional vector for $s=1,2$.

In the interaction picture associated with $H_c(t)$, the total Hamiltonian can be written as
\begin{eqnarray}
H(t)=H_{0}+H_{E}+U^{\dagger}_{c}(t)H_{\rm int}U_{c}(t),\label{Htot}
\end{eqnarray}
where $H_0$ is the Hamiltonian that performs the desired gate operation we want to protect and $H_{E}$ is the environmental Hamiltonian satisfying $U^{\dagger}_{c}(t)H_{E}U_{c}(t)=H_{E}$. We represent the
environment of each qubit as a thermal bath of harmonic
oscillators. In the case of a common environment for both qubits,
we consider
$H_{E}=\sum _{k} \omega_{k} {a_{k}}^{\dagger}a_{k}$,
where $\omega _{k}$ is the frequency of the $k$-th normal mode  of  the common environment,
and $a_{k}$ and ${a_{k}}^\dagger$ are the annihilation and creation operators, respectively.
In the case of two independent and identical environments, instead of 
the above we take $H_{E}=\sum_{s=1}^{2}\sum _{k} \omega_{k}
{a_{k}^{(s)}}^{\dagger}a_{k}^{(s)}$,  where $\omega _{k}$ is the
frequency of the $k$-th normal mode of the $s$-th qubit
environment, and $a_{k}^{(s)}$ and ${a_{k}^{(s)}}^\dagger$ are,
respectively, the corresponding annihilation and creation operators. The
frequency $\omega _{k}$ is the same for both independent and
identical environments. Accordingly, we take
$B_{m}^{(s)}  =  \sum_{k}\left(\lambda_{m}g_{k}^{\ast}a_{k}^{(s)}+
\lambda_{m}^{\ast}g_{k}a_{k}^{(s)\dagger}\right)$, 
where $g_{k}$ are coupling constants.

In a previous work, we have shown that it is possible to use a
continuously-applied external field to protect entangled states
from errors caused by the unavoidable interactions between the
qubit system and its environment \cite{fanchini07}. The question
naturally arises as to whether it is also possible to protect an
entangling operation. In the following, we show that, using the very same
external-field configuration of Ref. \cite{fanchini07}, we can prevent
errors from occurring during the application of a $\sqrt{\rm
SWAP}$ quantum gate to a two-qubit state under decoherence.

An ideal $\sqrt{\rm SWAP}$ gate is obtained by the Heisenberg coupling
between two qubits, whose dynamics is governed by the Hamiltonian
\begin{eqnarray}
H_{0}=J{\bm \sigma}^{(1)}\cdot{\bm \sigma}^{(2)},\label{H0}
\end{eqnarray}
where we use $\hbar =1$ throughout, $J$ is the exchange constant,
and, for $s=1,2$, ${\bm \sigma}^{(s)}={\bf
\hat{x}}\sigma_{x}^{(s)}+{\bf \hat{y}}\sigma_{y}^{(s)}+{\bf
\hat{z}}\sigma_{z}^{(s)}$, where $\sigma_{x}^{(s)}$,
$\sigma_{y}^{(s)}$, and $\sigma_{z}^{(s)}$ are the Pauli matrices
acting on qubit $s$. Remarkably, it has been shown that the
Heisenberg interaction alone is sufficient for universal quantum
computation, without the need of supplementary single-qubit
operations \cite{universalqc,kempe01,divincenzo00}. Thus, a protective scheme for
quantum gates based on this interaction, such as the $\sqrt{\rm SWAP}$ operation, is of fundamental
importance. Our results may find important applications in various experimental settings, such as double quantum dots \cite{petta05} and neutral atoms in optical lattices \cite{anderlini07}, where the exchange interaction between two-qubits can be realized.

It is possible to protect the considered quantum gate actualized by the Hamiltonian in Eq. (\ref{H0}) by the control Hamiltonian of the form
\begin{eqnarray}
H_{c}(t)={\bm \Omega}(t)\cdot\left( {\bm \sigma}^{(1)}+{\bm \sigma}^{(2)}\right).
\end{eqnarray}
In order for the evolution operator associated with the control Hamiltonian to satisfy Eq. (2) we must have \cite{fanchini07}:
\begin{eqnarray}
U_{c}(t)=U^{(2)}(t)U^{(1)}(t)=U^{(1)}(t)U^{(2)}(t),\label{Uc}
\end{eqnarray}
since ${\bm \sigma}^{(1)}$ and ${\bm \sigma}^{(2)}$ commute, where
\begin{eqnarray}
U^{(s)}(t)=\exp\left(-i\omega t n_{x}\sigma_{x}^{(s)}\right)
\exp\left(-i\omega t n_{z}\sigma_{z}^{(s)}\right),\label{Uk}
\end{eqnarray}
for $s=1,2$.

Equations (\ref{Uc}) and (\ref{Uk}) imply the following external field configuration:
\begin{eqnarray}
{\bm \Omega}(t)={\bf \hat{x}}n_{x}\omega +n_{z}\omega \left[{\bf
\hat{z}}  \cos\left( n_{x}\omega t \right)-{\bf \hat{y}}
\sin\left( n_{x}\omega t \right) \right].\label{Omega}
\end{eqnarray}
Here $\omega =2\pi/t_{c}$, $n_{x}$ and $n_{z}\ne n_{x}$ are
non-zero  integers, and $t_{c}$ is a constant. Such a field configuration is a combination
of a static field along the $x$ axis and a rotating
field in the $yz$ plane, and it is able to protect the evolution described by the Hamiltonian of Eq. (\ref{H0}) from  the effects of a
general class of errors. We can modify this field arrangement to be protective solely against a dephasing channel by setting $n_z=0$ in Eq. (\ref{Omega}). In this case we only have a static field along the $x$ axis, given by
${\bm \Omega}(t)={\bf \hat{x}}n_{x}\omega$, which is simpler than the field arrangement
given in Eq. (\ref{Omega}). Moreover, the field is supposed to be
spatially uniform in the neighborhood surrounding both qubits, since it is not necessary to address each qubit
independently.

Because Eq. (\ref{H0}) is a scalar product, it  is
invariant under rotations and
\begin{eqnarray}
U^{\dagger}_{c}(t)H_{0}U_{c}(t)=H_{0}.
\end{eqnarray}
This property of the Heisenberg interaction tremendously simplifies the quantum
operations executed under the protection by the CDD,
due to the fact that the intended  gate operation remains intact under the action of the control fields. Without this rotational invariance, we
would have to proceed as in Ref. \cite{fanchini022329} and
introduce an auxiliary rotating reference frame which complicates the
procedure. Furthermore, this invariance has another important property:
with the exact same field arrangement that preserves a quantum memory, we can
also protect the quantum-gate operation. Being able to protect the gate operation without the necessity of field reconfiguration is a tremendous simplification and certainly improves the prospects for experimental realization.

\section{Protecting the $\sqrt{\rm SWAP}$ gate} \label{sec}

\begin{figure} [h]
\includegraphics[width=.48\textwidth]{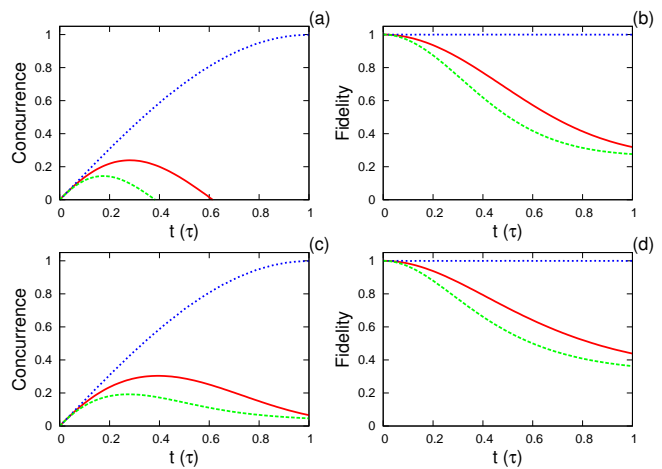}
\caption{(Color Online) \textit{Amplitude Damping plus Dephasing:} In figures (a)
and (b) we show, respectively, the concurrence and fidelity
for independent environments and, in figures (c) and (d), for the case of a
common environment. The dotted (blue) line represents the dynamics of
 a $\sqrt{\rm{SWAP}}$ quantum gate with protection. The solid (red) and dashed (green) lines represent the dynamics without protection for ohmic and super-ohmic
spectral densities, respectively.}\label{fig1}
\end{figure}

In order to illustrate our protective scheme, we take a Heisenberg interaction strength $J=\pi/8$ (c.f. Eq. (\ref{H0})) and a cut-off frequency for the spectral density, $J(\omega)=\eta\frac{\omega^s}{\omega_c^{s-1}}\exp(-\omega/\omega_{c})$, given by $\omega_c\tau=2\pi$, {as explained in the Appendix}, where $\tau=10^{-9}$s. We consider that the qubits interact with independent and common environments which are, in both cases, assumed to be at $T=0.2$K, with a coupling constant $\eta=1/20$, {as specified in the Appendix}. We investigate each environment model with ohmic (s=1) and super-ohmic (s=3) spectral densities. For the protected cases, we set the external field parameters as $n_x=28\pi/\tau$ and $n_z=14\pi/\tau$ (c.f. Eq. (\ref{Omega})).  We choose the initial state of the system to be $\rho(0)=|\!\!\uparrow\downarrow\rangle\langle\uparrow\downarrow\!\!|$. Note that the initial state is a product state and, in the absence of any decoherence, application of the $\sqrt{\rm SWAP}$ gate to this initial state will create a maximally entangled state.

\subsection{Amplitude damping and dephasing errors}

In Fig. (\ref{fig1}) we show the fidelity and concurrence of our two-qubit system, for the protected and unprotected cases, during the application of the $\sqrt{\rm SWAP}$ operation, when both amplitude damping and dephasing errors are present. We observe that, in the protected cases, the fidelity remains near unity during the whole time evolution, as opposed to the unprotected cases, where the fidelity decays for both ohmic and super-ohmic spectra, with a higher decay rate in the super-ohmic case. However, since the value of the concurrence is near unity at the end of the time evolution of the protected cases, we conclude that the entangling operation is successfully carried out. In fact, in the protected cases shown, higher values of fidelity and concurrence can be obtained for higher values of $n_x$ and $n_z$. However, the same is not true for the unprotected gate operation. For independent environments, the concurrence presents a peak, but then decays to zero with a lower maximum value and faster decay rate for the super-ohmic than for the ohmic spectrum. Finally, we see that the concurrence again presents a peak, followed by decay to a low but finite value, in the case of a common environment.

\subsection{Amplitude damping as a residual error}
\begin{figure} [h]
\includegraphics[width=.48\textwidth]{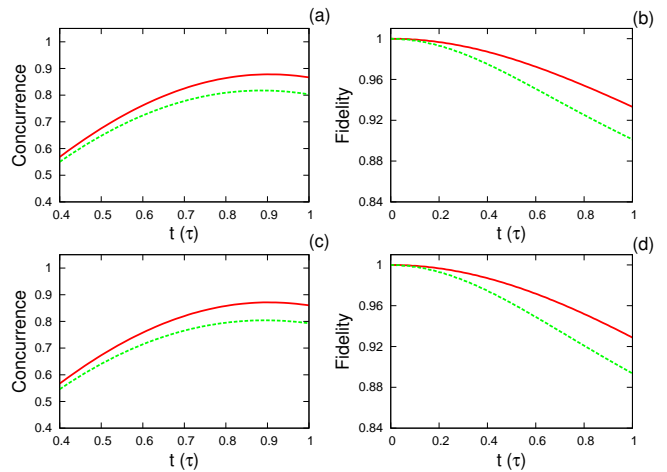}
\caption{(Color Online) \textit{Amplitude damping as a residual error channel:} In figures (a)
and (b) we show, respectively, the concurrence and fidelity
for independent environments and, in figures (c) and (d), for the case of a common environment. The solid (red) and dashed (green) lines
represent the dynamics of a $\sqrt{\rm{SWAP}}$ quantum gate for ohmic and super-ohmic environments, respectively.}\label{fig2}
\end{figure}

In Fig. (\ref{fig2}) we present the fidelity and concurrence of our two-qubit system, during the application of a $\sqrt{\rm{SWAP}}$ gate, protected only from dephasing, while the amplitude damping channel is left open as a residual error channel. We maintain the same external field configuration introduced in the beginning of this section, but set $n_z=0$, which makes our system vulnerable to the residual errors. Interestingly, we see that whether our two qubits interact independent or collectively with the environment has very little effect on the fidelity and concurrence. In both cases of interaction with the environment, the residual errors cause less damage to the system for the ohmic than for the super-ohmic spectral density.

\begin{figure} [h]
\includegraphics[width=.48\textwidth]{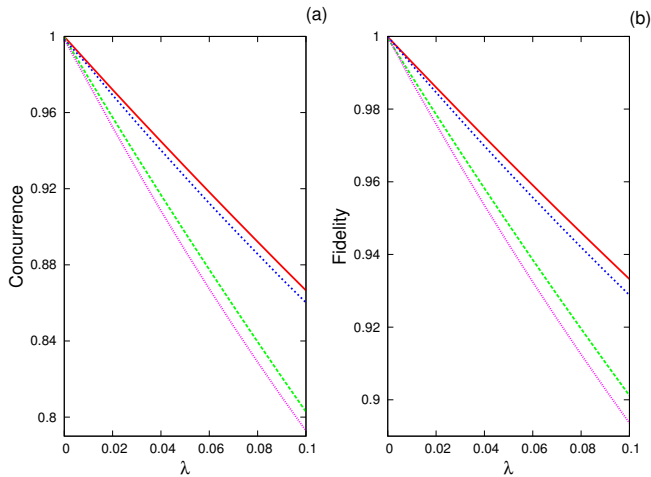}
\caption{(Color Online) The concurrence (a) and fidelity (b) at the end of the quantum gate operation, i.e., at $t=\tau$,  as functions of the coupling constant, $\lambda$, of the residual error channel. The solid (red) line represents the case of independent environments with ohmic spectra, the dotted (blue) line represents the case of a common environment with an ohmic spectrum, the dashed (green) line represents the case of independent environments with super-ohmic spectra, and the purple (closely-dotted) line represents the case of a common environment with a super-ohmic spectrum.}\label{fig3}
\end{figure}

In Fig. (\ref{fig3}) we show the concurrence and fidelity at the end of the quantum gate operation, i.e., at $t=\tau$, as functions of the amplitude-damping coupling constant $\lambda$ (see Eqs. (\ref{collective}) and (\ref{independent})). For all possible cases of environmental spectral density and interaction, the concurrence and fidelity decrease linearly as functions of the coupling constant. While the case of a common environment with an ohmic spectral density presents the highest decay rate, the case of independent environments with ohmic spectra shows the lowest decay rates, as functions of $\lambda$. However, it is important to note that, regarding these functions, there is little difference between the case of a common environment and that of two independent environments. A plausible argument to explain this behaviour would be to recall the fact that in our case qubit-qubit coupling is stronger than qubits' coupling to the environment. Therefore, any environmental effect on one of the qubits is quickly felt by the other qubit for both common and independent environments. 

\section{Conclusion}

In summary, we have presented a CDD strategy to protect, from general and residual errors, a $\sqrt{{\rm SWAP}}$ quantum gate, which is an entangling operation realizable using the Heisenberg exchange interation. The gate operation is applied to a two-qubit system and the errors are introduced by amplitude-damping and dephasing channels, resulting from interactions with bosonic environments. We consider both common and independent environments, together with ohmic and super-ohmic environmental spectral densities. We quantify the success of protection by looking at the fidelity and concurrence of the system during the time of gate operation. When both error mechanisms are present and no protection is supplied, we observe that the case of two independent environments with super-ohmic spectral densities is the most harmful. However, we have shown that our protection scheme works very well against these errors, keeping the system at high fidelity and letting the entangling gate operation perform successfully. We have also considered a residual error setting, where only dephasing errors are protected, while amplitude-damping errors are allowed to affect the system. In this case, we have seen that independent or common environmental interactions show little difference and the case of a super-ohmic spectral density continues to be more harmful to the gate operation as in the previous case. Furthermore, our CDD scheme uses the same external field configuration in the dynamic as in the static case, which is an important property for experimental applications.

{Finally, we would like to say a few words about the dependence of the concurrence and fidelity on the spectral function of separable and common environments. To start with, let us address the issue of separability.}

{As a matter of fact, this dependence only shows up in the rate of change of those quantities. Separable environments seem to destroy them faster than the common ones. Although we are not going to provide any detailed explanation for this fact, we can argue that quantum coherence properties between two qubits are more likely to be preserved by the presence of a common environment since it could mediate, at least in some  cases, an indirect effective coupling between them. In other words, in common environments there might be the possibility of existence of some cooperative effect on top of the deleterious effects present in any coupling to general environments. On the other hand, separable environments would always act independently on each qubit giving them very little chance to preserve or develop any quantum mechanical coherence.}

{As for the spectral function dependence, the stronger effect of superohmic environments as compared to that of ohmic environments can be simply understood in terms of time scales.}
{It is a very simple matter for the reader to convince him or herself that the general spectral function $J(\omega)$ defined in the beginning of section \ref{sec} has a maximum at $\omega_{m}^{(s)}=s\omega_{c}$ whose value is $J_{max}^{(s)}\equiv J(\omega_{m}^{(s)})=\eta\omega_{c}(s/e)^{s}$. Therefore, we see that for superohmic environments (with $s=3$) $J(\omega)$ is peaked at $3\omega_{c}$ with its maximum value given by $J_{m}^{(3)} \approx 1.1 \eta\omega_{c}$ whereas these values are, respectively, $\omega_{c}$ and $J_{m}^{(1)} \approx 0.4 \eta\omega_{c}$ for ohmic environments. Another important point to be observed is that since we are interested in protection schemes, and consequently testing our system for times in the interval $0 < t < \tau = 2\pi/\omega_{c}$, the relevant frequency range for our analysis is $\omega > \omega_{c}/2\pi$.}

{Therefore we see that as time evolves from $t=0$, the deleterious effects of the superohmic environment start to take place earlier than those of the ohmic environment because  the spectral weight of the former is more pronounced at a higher frequency than that of the latter. In other words, it is the high frequency behavior of the spectral function which dominates any phenomena at this time scale.}

{At this point one should naturally argue that our reasoning does not make much sense for the present problem because the form established for the spectral function $J(\omega)$ is actually appropriate for dealing with phenomena at very long time scales ($t\gg 1/\omega_{c}$) which are  dominated by the low frequency behavior of $J(\omega)$. The cutoff frequency $\omega_{c}$ is a characteristic frequency of the environment which fixes the time scale of the problem. Nevertheless, we can still sustain our results if we take them as an indication that no matter what bath we have, it is its high frequency behavior that matters in this case. This means that if one is really interested in a more quantitative analysis of the problem, a more detailed account of the environment's high frequency behavior is in order.}

\section{Acknowledgements}

This work has been partly supported by ``Funda\c{c}{\~a}o de Amparo \`a
Pesquisa do Estado de S{\~a}o Paulo (FAPESP)'', Brazil, project numbers 12/50464-0 and 2014/21792-5, and by ``Conselho Nacional de
Desenvolvimento  Cient\'ifico e Tecnol\'ogico (CNPq)'', Brazil, through
 the ``Instituto Nacional de Ci{\^e}ncia e Tecnologia em
Informa\c{c}{\~a}o Qu{\^a}ntica (INCT-IQ)''.

\appendix*

\section{Solution of the master equation}
Starting from $H(t)$, the Hamiltonian in the interaction picture is written as
\begin{eqnarray}
H_{I}(t)=\sum _{s=1} ^{2}\sum _{m=1} ^{3}\sum_{n =1}^{3}R_{m,n}(t)E_{m}^{(s)}(t)
\widetilde{\sigma }_{n}^{(s)}(t),\label{HI}
\end{eqnarray}
where $\sigma_{1}^{(s)}\equiv \sigma_{x}^{(s)}$,
$\sigma_{2}^{(s)}\equiv \sigma_{y}^{(s)}$, $\sigma_{3}^{(s)}\equiv
\sigma_{z}^{(s)}$, $\widetilde{\sigma }_{n}^{(s)}(t)=
U^{\dagger}_{0} (t)\sigma _{n}^{(s)}U_{0} (t)$, for $s=1,2$ and
$n=1,2,3$, with $U_{0}(t)=\exp(-iH_{0}t)$. We have used {Eq. (\ref{Hint}) and defined the operators}
$E_{m}^{(s)}(t)=U^{\dagger}_{E}(t)B_{m}^{(s)}U_{E}(t)$, for
$s=1,2$ and $m=1,2,3$, with $U_{E}(t)=\exp(-iH_{E}t)$. The
quantities $U^{\dagger}_{c} (t)\sigma_{m}^{(s)}U_{c} (t)=\sum_{n
=1}^{3} R_{m,n}(t) \sigma_{n}^{(s)}$, for $s=1,2$ and $m=1,2,3$,
are rotations of $\sigma_{m}^{(s)}$, whose matrix elements,
$R_{m,n}(t)$, are real functions of time. We proceed as  in Ref.
\cite{fanchini07} and assume that the absolute temperature is the
same in the surroundings of both qubits and these qubits, as well
as their respective environments, are identical. We then write
down the master equation for the two-qubit reduced density matrix,
$\rho_{I}(t)$, in the Born approximation:
\begin{widetext}
\begin{eqnarray}
\frac{d\rho_{I}(t)}{dt} & = & \sum_{s,s^{\prime}=1}^{2}
\sum_{n,n^{\prime}=1}^{3}\int_{0}^{t}dt^{\prime}\,
\left\{\mathcal{D}_{n,n^{\prime}}^{(s,s^{\prime})}(t,t^{\prime})
[\widetilde{\sigma}_{n}^{(s)}(t),\rho_{I}(t)\widetilde{\sigma}_{n^{\prime}}^{(s^{\prime})}(t^{\prime})]
+[\mathcal{D}_{n,n^{\prime}}^{(s,s^{\prime})}(t,t^{\prime})]^{*}
[\widetilde{\sigma}_{n^{\prime}}^{(s^{\prime})}(t^{\prime})\rho_{I}(t),
\widetilde{\sigma}_{n}^{(s)}(t)]\right\},\label{master}
\end{eqnarray}
\end{widetext}
where we have {defined} the coefficients
\begin{eqnarray*}
\mathcal{D}_{n,n^{\prime}}^{(s,s^{\prime})}(t,t^{\prime}) & = &
\sum_{m=1}^{3}
\sum_{m^{\prime}=1}^{3}R_{m,n}(t)R_{m^{\prime},n^{\prime}}(t^{\prime})
C_{m,m^{\prime}}^{(s,s^{\prime})}(t,t^{\prime}),\end{eqnarray*}
for $n,n^{\prime}=1,2,3$ and $s,s^{\prime}=1,2$,
and\begin{eqnarray*}
C_{m,m^{\prime}}^{(s,s^{\prime})}(t,t^{\prime}) & = &
\mathrm{Tr}_{E} \left\{
E_{m}^{(s)}(t)\rho_{E}E_{m^{\prime}}^{(s^{\prime})}(t^{\prime})\right\}
,\end{eqnarray*} for $m,m^{\prime}=1,2,3$ and  $s,s^{\prime}=1,2$.
$C_{m,m^{\prime}}^{(s,s^{\prime})}(t,t^{\prime})$ is the
correlation function between components $m$ and $m^{\prime}$ of
environmental operators calculated at the same qubit position, as
explained in Ref. \cite{fanchini07}. Here, $\mathrm{Tr}_{E}$
denotes the trace over the environmental degrees of freedom. The
operators $\widetilde{\sigma}_{n}^{(s)}(t)$, for $s=1,2$ and
$n=1,2,3$, can be explicitly obtained as the components of the
following vector relations:
\begin{eqnarray}
U^{\dagger}_{0} (t){\bm \sigma }^{(1)}U_{0} (t)=a(t){\bm \sigma }^{(1)}+b(t){\bm \sigma }^{(2)}\nonumber \\
-c(t)({\bm \sigma }^{(1)}\times {\bm \sigma }^{(2)}),
\end{eqnarray}
and
\begin{eqnarray}
U^{\dagger}_{0} (t){\bm \sigma }^{(2)}U_{0} (t)=a(t){\bm \sigma }^{(2)}+b(t){\bm \sigma }^{(1)}\nonumber \\
-c(t)({\bm \sigma }^{(2)}\times {\bm \sigma }^{(1)}),
\end{eqnarray}
where $a(t)=[1+\cos (4Jt)]/2$, $b(t)=[1-\cos (4Jt)]/2$, and
$c(t)=\sin (4Jt)/2$.  The environmental density matrix, $\rho
_{E}$, is taken as the one for a canonical ensemble constituting a
thermal bath, that is,
$\rho _{E}=\frac{1}{Z}\exp(-\beta H_{E})$,
where $Z$ is the partition function, $Z={\mathrm
Tr}_{E}\left[\exp(-\beta H_{E})\right]$.  Here, $\beta =1/k_{B}T$,
$k_{B}$ is the Boltzmann constant, and $T$ is the absolute
temperature of the environment.

We can also write the correlation function as\begin{eqnarray*}
C_{m,m^{\prime}}^{(s,s^{\prime})}(t,t^{\prime}) & = & \Gamma^{(s,s^{\prime})}\mathrm{Tr}_{E}
\left\{ E_{m}^{(s)}(t)\rho_{E}E_{m^{\prime}}^{(s)}(t^{\prime})\right\} ,\end{eqnarray*}
where $\Gamma^{(s,s^{\prime})}=1$ for the case of a single, common
environment, in which case the environmental operators $E_{m}^{(s)}(t)$
are independent of $s$, and $\Gamma^{(s,s^{\prime})}=\delta_{s,s^{\prime}}$
for the case of two identical, uncorrelated environments. Since we
have
\begin{eqnarray*}
E_{m}^{(s)}(t) & = & \sum_{k}\left[\lambda_{m}g_{k}^{\ast}a_{k}^{(s)}e^{-i\omega_{k}t}+
\lambda_{m}^{\ast}g_{k}a_{k}^{(s)\dagger}e^{+i\omega_{k}t}\right]\end{eqnarray*}
and, therefore,

\begin{eqnarray*}
\mathrm{Tr}_{E}\!\!\left\{\! E_{m}^{(s)}(t)\rho_{E}E_{m^{\prime}}^{(s)}(t^{\prime})\!\right\}\! &=& \!
\lambda_{m}\lambda_{m^{\prime}}^{\ast}\sum_{k}\left|g_{k}\right|^{2}n_{k} e^{-i\omega_{k}(t-t^{\prime})}\\
&&\hspace{-1.05cm}+\,\lambda_{m}^{\ast}\lambda_{m^{\prime}}\sum_{k}\left|g_{k}\right|^{2}(1+n_{k})
e^{i\omega_{k}(t-t^{\prime})},
\end{eqnarray*}
where $ n_{k}=1/[\exp(\beta\omega_{k})-1]$, we obtain
\begin{widetext}
\begin{eqnarray*}
\frac{d\rho_{I}(t)}{dt} & = & \sum_{s,s^{\prime}=1}^{2}
\int_{0}^{t}dt^{\prime}\,\mathcal{T}_{1}^{(s,s^{\prime})}(t-t^{\prime})
[\mathcal{R}^{(s)}(t),\rho_{I}(t)[\mathcal{R}^{(s^{\prime})}
(t^{\prime})]^{\dagger}]+\sum_{s,s^{\prime}=1}^{2}\int_{0}^{t}dt^{\prime}\,
\mathcal{T}_{2}^{(s,s^{\prime})}(t-t^{\prime})
[[\mathcal{R}^{(s)}(t)]^{\dagger},\rho_{I}(t)\mathcal{R}^{(s^{\prime})}(t^{\prime})]\\
 & + & \sum_{s,s^{\prime}=1}^{2}\int_{0}^{t}dt^{\prime}\,
 [\mathcal{T}_{1}^{(s,s^{\prime})}(t-t^{\prime})]^{*}
 [\mathcal{R}^{(s^{\prime})}(t^{\prime})\rho_{I}(t),[\mathcal{R}^{(s)}(t)]^{\dagger}]+
 \sum_{s,s^{\prime}=1}^{2}\int_{0}^{t}dt^{\prime}\,
 [\mathcal{T}_{2}^{(s,s^{\prime})}(t-t^{\prime})]^{*}
 [[\mathcal{R}^{(s^{\prime})}(t^{\prime})]^{\dagger}\rho_{I}(t),\mathcal{R}^{(s)}(t)]\end{eqnarray*}
\end{widetext}
where
\begin{eqnarray*}
\mathcal{R}^{(s)}(t) & = & \sum_{m=1}^{3}\sum_{n=1}^{3}\lambda_{m}R_{m,n}(t)\widetilde{\sigma}_{n}^{(s)}(t),
\end{eqnarray*}
\begin{eqnarray*}
\mathcal{T}_{1}^{(s,s^{\prime})}(t) & = & \Gamma^{(s,s^{\prime})}\sum_{k}\left|g_{k}\right|^{2}n_{k}\exp(-i\omega_{k}t),
\end{eqnarray*}
and
\begin{eqnarray*}
\mathcal{T}_{2}^{(s,s^{\prime})}(t) & = & \Gamma^{(s,s^{\prime})}\sum_{k}\left|g_{k}\right|^{2}(1+n_{k})
\exp(i\omega_{k}t).
\end{eqnarray*}

In the limit the number of environmental normal modes per
unit frequency becomes {very large}, we define a spectral density as $J(\omega)=\sum_{k}|g_{k}|^{2}
\delta(\omega-\omega_{k})$,
with $\omega\in\left[0,\infty\right)$, and interpret the summations
in $\mathcal{T}_{1}^{(s,s^{\prime})}(t)$ and $\mathcal{T}_{2}^{(s,s^{\prime})}(t)$
as integrals over $\omega$:\begin{eqnarray*}
\mathcal{T}_{1}^{(s,s^{\prime})}(t) & = & \Gamma^{(s,s^{\prime})}
\int_{0}^{\infty}d\omega J(\omega)\frac{\exp(-i\omega t)}{\exp(\beta\omega)-1},\end{eqnarray*}
 and \begin{eqnarray*}
\mathcal{T}_{2}^{(s,s^{\prime})}(t) & = & \left[\mathcal{T}_{1}^{(s,s^{\prime})}(t)\right]^{\ast}+
\Gamma^{(s,s^{\prime})}\int_{0}^{\infty}d\omega J(\omega)\exp(i\omega t),\end{eqnarray*}
for $s,s^{\prime}=1,2$. Here we assume
a cutoff frequency $\omega_{c}$, and write {$J(\omega)=\eta (\omega^s/\omega_c^{s-1})\exp(-\omega/\omega_{c})$} ,
where $\eta$ is a dimensionless coupling constant, and the values of $s$ specify the two kinds of environmental spectral density we treat: ohmic ($s=1$)  and super-ohmic ($s>1$).

\end{document}